\documentclass[preprint,prl]{revtex4}
\usepackage{setspace}
\usepackage{graphicx}
\usepackage{color}
\usepackage{epsfig}
\usepackage{subfigure}
\usepackage{natbib}

\begin{document}

\title{First-principles theory of coloration of WO$_3$ upon charge insertion}

\author{Yu Xue}

\affiliation{Department of Physics,
University at Buffalo,
State University of New York,
Buffalo, NY 14260, USA
}

\author{Yong Zhang}
\affiliation{National Renewable Energy Laboratory, 1617 Cole Boulevard, Golden, Colorado 80401, USA}
\author{Peihong Zhang}

\affiliation{Department of Physics,
University at Buffalo,
State University of New York,
Buffalo, NY 14260, USA}

\date{\today}

\begin{abstract}
We report first-principles investigations of the coloration of WO$_3$ upon charge insertion, 
using sodium tungsten bronze (Na$_x$WO$_3$) as a model system. Our results explain  
well the systematic color change of Na$_x$WO$_3$ from dark blue to violet, red-orange, and finally to
golden-yellow as sodium concentration $x$ increases from 0.3 to unity.
Proper accounts for both the interband and the intraband contributions
to the optical response are found to be very important for a detailed understanding of the
coloration mechanism in this system.
\end{abstract}

\maketitle

Electrochromic materials~\cite{handbook,APPL_PHYS_A_89_29_2007} exhibit reversible
and persistent change of the optical properties, hence their color, upon applying an 
electrical pulse that injects both electrons and compensating ions into the material. 
Potential applications of electrochromic materials range from information display, 
light shutters, to energy efficient smart windows~\cite{handbook,APPL_PHYS_A_89_29_2007}.  
Tungsten trioxide~\cite{Deb_1969,Deb_1973} is one of the most extensively studied electrochromic 
materials due to its superior coloration efficiency, short response time and reversibility~\cite{handbook}.
Enhanced electrochromic properties in WO$_3$ nanowires have been reported recently~\cite{nanowires1,nanowires2}.
Despite much research effort, a first-principles theory for the coloration mechanism in
this material has not emerged. Here we report density functional theory (DFT)~\cite{DFT}
investigations of the coloration mechanism of WO$_3$ upon charge insertion. 
Our results explains very well the systematic change in color~\cite{handbook,JB_Goodenough,S_Raj} 
of Na$_x$WO$_3$ from dark blue to metallic golden-yellow with increasing sodium concentration. 
We find that proper accounts for the free-carriers contribution to the optical response are critical 
for a quantitative understanding of the coloration mechanism in this system.

Undoped stoichiometric WO$_3$ is a transparent semiconductor. 
Double injection of electrons and compensating ions (e.g., H$^+$, Li$^+$, or Na$^+$) by
an electric pulse induces a dark blue color in the electrochromic WO$_3$ film.
Despite intense research over the last three decades, the mechanism of electrochromism remains 
controversial~\cite{handbook,Faughnan_1975,APL_27_646_1975,S_Hashimoto_J_APPL_PHYS_69_933_1991,
Monk_1999,PRB_68_233204_2003}.
Interestingly, sodium tungsten bronze (Na$_x$WO$_3$) also shows various bluish colors at low
sodium concentrations. Therefore, it is likely that both systems share the same fundamental 
coloration mechanism~\cite{APL_27_646_1975},
and the formation of tungsten bronze might be responsible for the electrochromism in WO$_3$.
Furthermore, the color of Na$_x$WO$_3$ changes from dark blue to violet, red-orange, and finally to 
golden-yellow as sodium concentration $x$ increases 
from 0.3 to unity~\cite{handbook,JB_Goodenough,S_Raj}.
A consistent theory should be able to explain the color of Na$_x$WO$_3$ for a wide range of 
Na concentration. 
Unfortunately, although the electronic and structural properties 
of Na$_x$WO$_3$ has been studied 
extensively~\cite{JB_Goodenough,S_Raj,JF_Owen_PRB_18_3827_1978,MA_Langel_PRB_23_1584_1981,
JH_Davies_PRL_57_475_1986,RG_Egdell_CPL_85_140_1982,M_Kielwein_PRB_51_10320_1995,PRB_54_2436_1996,PRB_70_165110_2004,PRB_72_075109_2005},
a quantitative understanding of its vivid color change with varying Na content is still lacking.

{\it Materials color from first principles ---}
The apparent color of a material is intimately related to its optical absorption spectra. 
With the advance of modern electronic structure methods, various optical properties can 
be routinely calculated nowadays. However, to the best of our knowledge, there have been 
no reports of calculating the apparent color of materials from first-principles.   
The complicated physiology of human color perception makes calculating
the apparent color of a material rather difficult. 
Fortunately, the color space~\cite{CIE} defined by the International Commission of Illumination (CIE), 
the CIE 1931 XYZ color space, provides 
a mathematical foundation for a quantitative description of colors. 

\begin{figure}
\includegraphics[width=3.0in,angle=0]{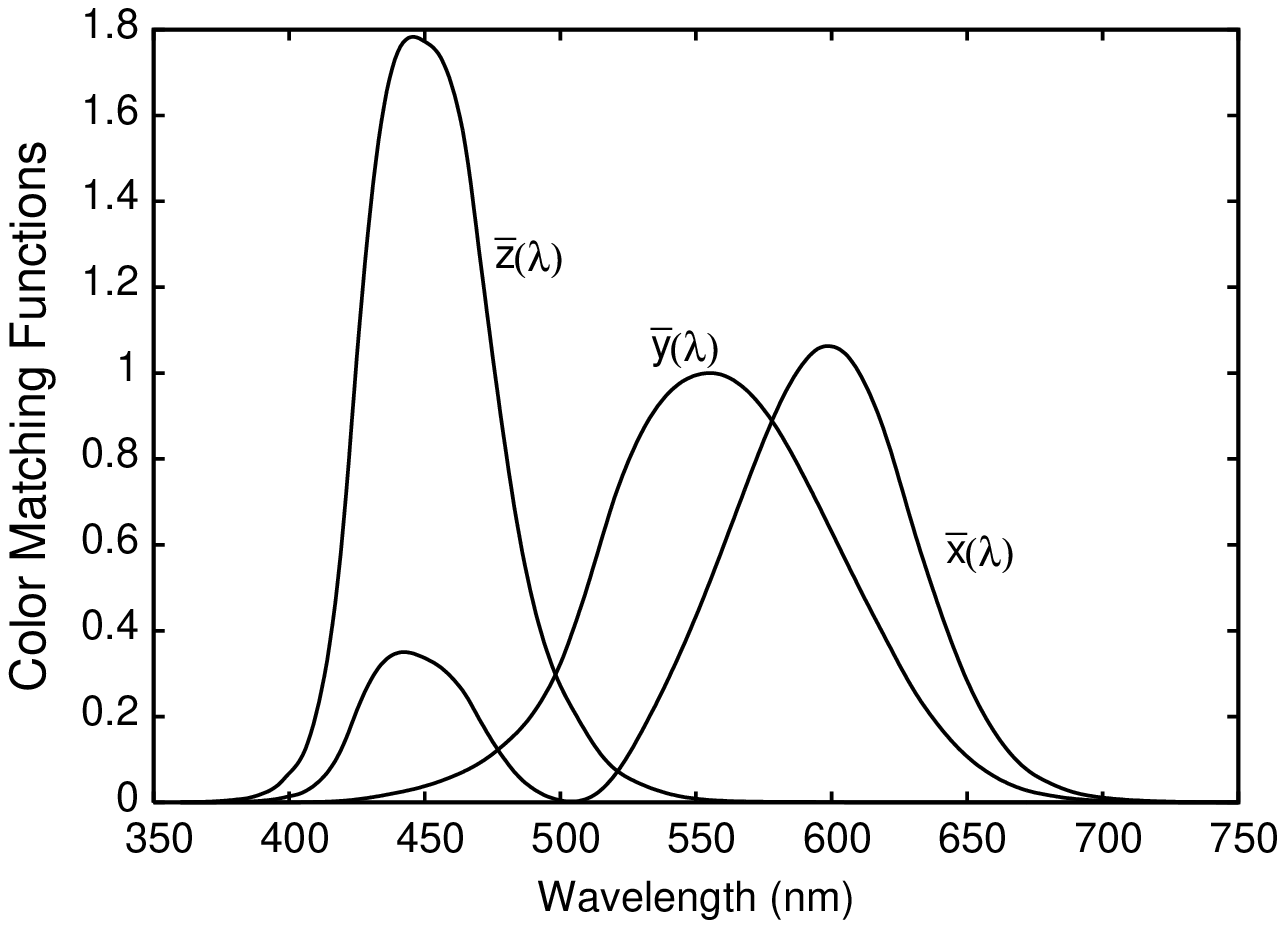}
\caption{The CIE XYZ color matching functions.
The CIE defined three color-matching functions, known
as $\bar{x}(\lambda)$, $\bar{y}(\lambda)$, and $\bar{z}(\lambda)$. These
functions can be considered as the spectral sensitivity functions
of three linear light detectors that give the CIE XYZ tristimulus
values.}  
\label{CIE}
\end{figure}

The CIE XYZ color space defines three color matching functions, 
$\bar{x}(\lambda)$, $\bar{y}(\lambda)$, and $\bar{z}(\lambda)$ as shown in Fig.\ \ref{CIE}.
The XYZ tristimulus values for an object illuminated by a light source with 
a spectral power distribution $I(\lambda)$ are then given by
\begin{eqnarray}
\nonumber X=&\int{I(\lambda)R(\lambda)\bar{x}(\lambda)d\lambda}; \\
\nonumber Y=&\int I(\lambda)R(\lambda)\bar{y}(\lambda)d\lambda; \\
          Z=&\int I(\lambda)R(\lambda)\bar{z}(\lambda)d\lambda, 
\label{e1}
\end{eqnarray}
\noindent where $R(\lambda)$ [or $R(\omega)$] is the wavelength (frequency) dependent reflectivity of the object.
The CIE XYZ color space is related to the well-known CIE RGB color space by the
following linear transformation:
\begin{equation}
\left[
\begin{array}{c}
X \\
Y \\
Z
\end{array}
\right]
=\frac{1}{0.17697}
\left[
\begin{array}{ccc}
0.49000 & 0.31000 & 0.20000 \\
0.17697 & 0.81240 & 0.01063 \\
0.00000 & 0.01000 & 0.99000 
\end{array}
\right]
\left[
\begin{array}{c}
R \\
G \\
B
\end{array}
\right].
\label{e2}
\end{equation}
\noindent The reflective color and the relative brightness of a material can be constructed
once we obtain the tristimulus values X, Y, and Z, or alternatively, R, G, and B.

For an ideal semi-infinite crystal, the normal incidence reflectivity is related to the 
complex index of refraction $\tilde{n}$ and the macroscopic complex dielectric function 
$\varepsilon(\omega)=\varepsilon_1(\omega)+i\varepsilon_2(\omega)$ via
\begin{eqnarray}
R(\omega)=|\frac{\tilde{n}(\omega)-1}{\tilde{n}(\omega)+1}|^2; \hspace{0.4cm}  \tilde{n}(\omega)=\varepsilon^{1/2}(\omega).
\label{e3}
\end{eqnarray}
\noindent The imaginary part of the dielectric function $\varepsilon_2(\omega)$ 
due to interband transitions can be calculated using DFT-based 
first-principles electronic structure techniques: 
\begin{equation}
\varepsilon^{\mathrm{inter}}_2(\omega)=\frac{16\pi^2}{\omega^2}\sum_{vc\vec{k}}
|\vec{\lambda}\cdot\langle v\vec{k}|\vec{v}|c\vec{k}\rangle|^2
\delta[\omega-(\epsilon_{c\vec{k}}-\epsilon_{v\vec{k}})],
\label{e4}
\end{equation}
\noindent where $\vec{\lambda}$ is the polarization of light, $\vec{v}$ is the velocity operator, 
$|c\vec{k}\rangle$ ($|v\vec{k}\rangle$) denotes unoccupied conduction states (occupied valence states). 
The real part of the dielectric function is related to the imaginary part through the Kramers-Kronig relations.
In undoped semiconductors, only the interband
transitions contribute to the optical response. In metals or doped semiconductors, however, 
the free-carriers contributions to the optical 
response are very important. These contributions can be conveniently modeled on the basis of the Drude theory:
\begin{equation}
\varepsilon^{\mathrm{intra}}_2(\omega)=\frac{\gamma\omega^2_p}{\omega(\gamma^2+\omega^2)}, \hspace{0.4cm}
\omega^2_p=4\pi n/m^*. 
\label{e5}
\end{equation}
\noindent 
In the above equations, $n$, $m^*$, and $1/\gamma$ are, respectively, the density, 
optical effective mass, and relaxation time of the free carriers.
Combining both the interband and intraband contributions thus gives
the imaginary part of the total electronic dielectric function $\epsilon^{\mathrm{total}}_2=
\epsilon^{\mathrm{inter}}_2+
\epsilon^{\mathrm{intra}}_2$.

{\it Electronic structure and optical properties of Na$_{1.0}$WO$_3$ ---}
Tungsten trioxide comprises of corner-sharing WO$_6$ octahedra~\cite{handbook,Salje}. 
Several metal ions can be incorporated into the structure to form tungsten bronzes. Among them 
sodium tungsten bronze (Na$_x$WO$_3$) is the most extensively studied
system owing to its unusual optical properties and dramatic changes in color\cite{handbook,JB_Goodenough,S_Raj}
with increasing $x$. Na$_x$WO$_3$ assumes a simple cubic structure for Na contents $x\ge 0.4$ and undergoes a 
sequence of structural changes with decreasing $x$~\cite{Ribnick,Brown_Banks}. 
The cubic phase can be synthesized under nonequilibrium conditions down 
to $x\sim$ 0.2~\cite{M_Kielwein_PRB_51_10320_1995,J_CHEM_PHYS_36_87_1962}.
Our study covers Na concentration $0.3\le x \le 1.0$. For  simplicity, we assume a cubic structure 
for all Na concentrations studied.
Our calculations are based on DFT within the local density 
approximation (LDA)~\cite{LDA}. In order to avoid inconvenient (i.e., large) unit cells for systems with
fractional $x$, we adopt the virtual crystal approximation\cite{VCA1,VCA2} in which fractional 
Na compositions are modeled by pseudoatoms with appropriate nuclear charges and valence electrons. 
We use the experimental lattice constants of Na$_x$WO$_3$, which can be conveniently 
expressed by the Vegard's law $a(x)=3.7845+0.0821x$~\cite{Brown_Banks}. The Brillouin zone 
summation in Eqn.\ \ref{e4} is carried out on a dense $32\times 32\times 32$ uniform grid with an energy broadening of 0.3 eV. 

\begin{figure}
\begin{center}
\begin{tabular}{c}
\resizebox{3.0in}{!}{\includegraphics[angle=0]{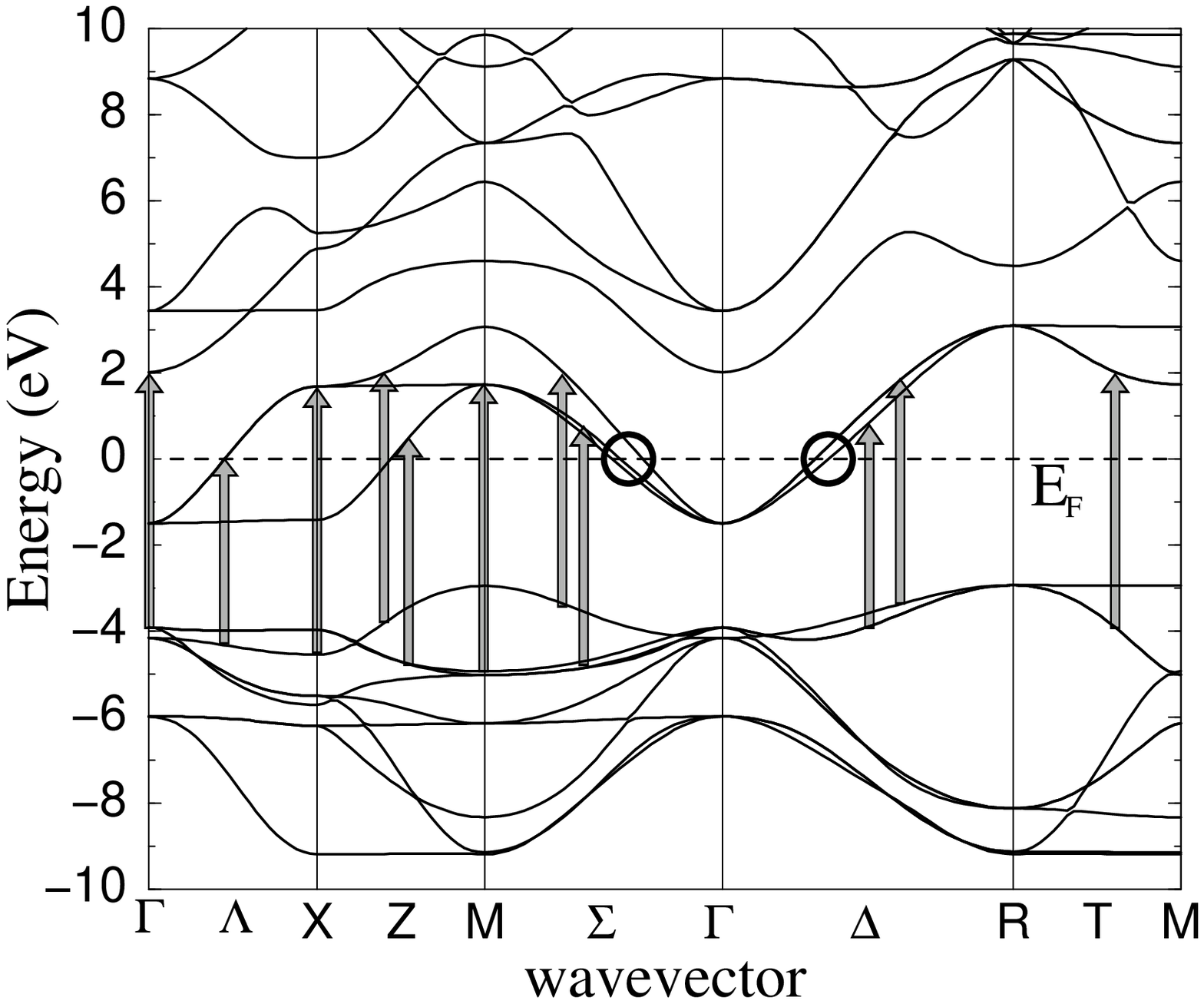}} \\
\resizebox{3.0in}{!}{\includegraphics[angle=-90]{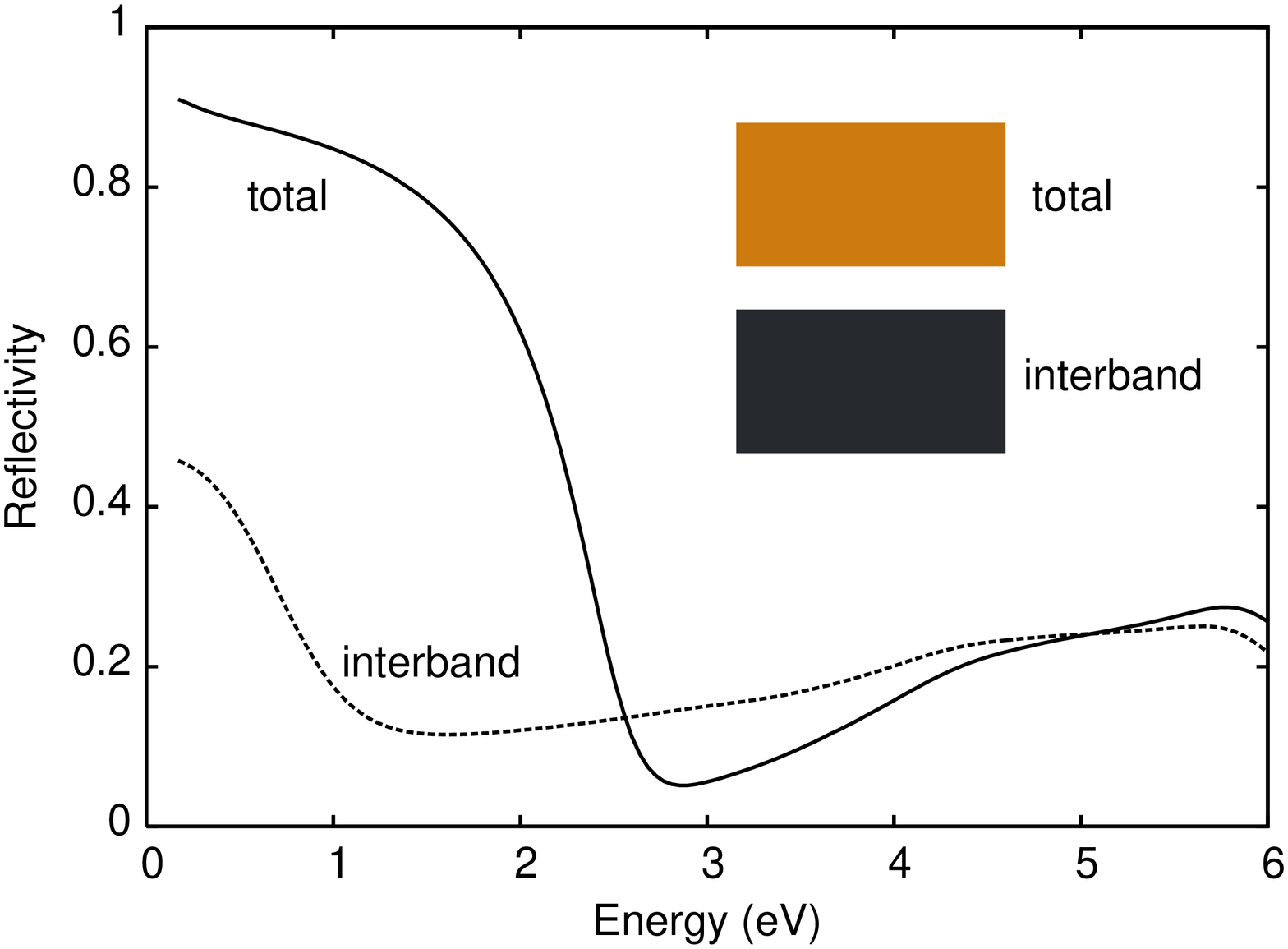}} \\
\resizebox{3.0in}{!}{\includegraphics[angle=0]{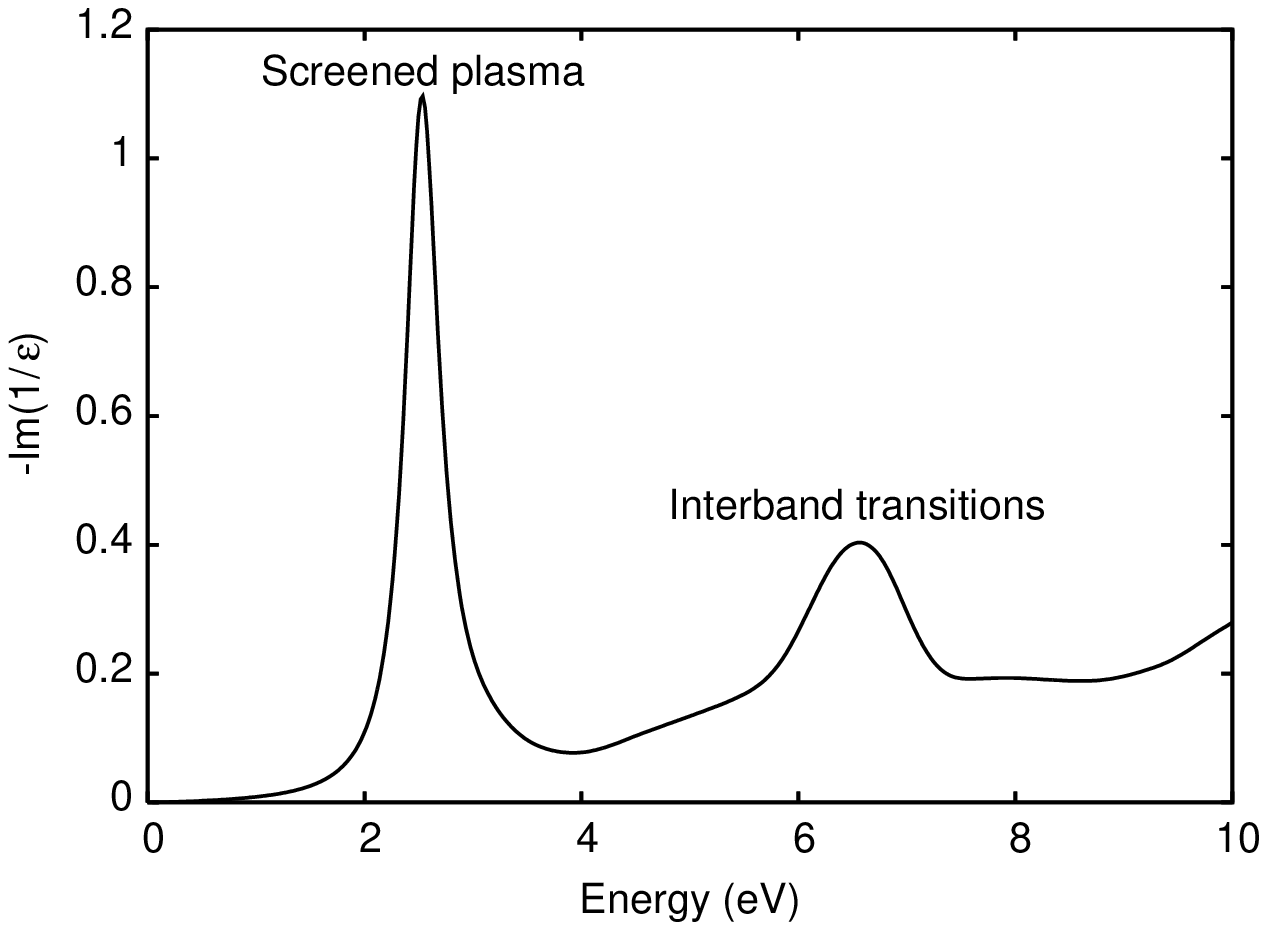}} \\
\end{tabular}
\caption{Band structure (top), reflectivity (middle) and loss function of Na$_{1.0}$WO$_3$ (bottom). 
Some strong interband transitions are indicated by shaded arrows.  Strong inter-subband transitions 
are indicated by circles. The reflectivity spectra are calculated with and without
the free-carriers contributions. Inset shows the colors of Na$_{1.0}$WO$_3$ before and after taking into
account the free-carriers response.
}
\label{bs_reflectivity}
\end{center}
\end{figure}

Figure \ref{bs_reflectivity} (a) shows the band structure of Na$_{1.0}$WO$_3$, the ending system of the series studied. 
As mentioned previously, Na$_{1.0}$WO$_3$ displays a distinct golden-yellow color. The electronic band structure suggests, 
however, that the strong interband transitions mainly occur across an
energy window of $4\sim 7$ eV. This falls well above the visible energy window and
cannot explain its golden-yellow appearance.
There are also inter-subband transitions as indicated by circles in 
Fig.\ \ref{bs_reflectivity} (a). However, the energies of these transitions are very low (less than 1.0 eV)
 and cannot be responsible for the optical response in the visible range neither. 
Therefore, a proper account for the free carriers (intra-band) contributions is critical for a 
quantitative understanding of the optical properties of this system. 

Figure \ref{bs_reflectivity} (b) shows the calculated reflectivity spectra $R(\omega)$ of 
Na$_{1.0}$WO$_3$, with and without including the 
free-carriers response.
The unscreened plasma frequency in Eqn.\ \ref{e5} is calculated using the  measured effective 
mass~\cite{JF_Owen_PRB_18_3827_1978}. It is clear that the reflectivity in 
the visible range (1.5 $\sim$ 3.0 eV) is dominated by the free carriers contribution (which
is renormalized by the interband transitions as explained below). 
The spectra is fairly uniform and featureless for photon energy $1.5 \le E \le 3.0$ eV 
if the free carriers plasma response is not included. 
 As a result, Na$_{1.0}$WO$_3$ 
would appear dark gray (more precisely, slightly bluish dark gray as shown in the inset of
Fig.\ \ref{bs_reflectivity} (b) without taking into account the free-carriers contributions.
These contributions significantly enhance the
reflectivity below 2.5 eV and suppresses the reflectivity around 3.0 eV, 
resulting in the golden-yellow appearance of Na$_{1.0}$WO$_3$ as shown in the inset of 
Fig.\ \ref{bs_reflectivity} (b). 
The onset of the free-carriers enhanced reflectivity corresponds
to the renormalized (screened) plasma frequency $\tilde{\omega}_p$,
defined by $\mathrm{Re}[\epsilon^{-1}(\tilde{\omega}_p)]=0$. The renormalized
plasma frequency is related to the bare value (defined in Eqn.\ \ref{e5})
by $\tilde{\omega}_p\approx [\omega_p^2/\epsilon_1^{inter}(0)]^{1/2}$, which
can be measured in electron energy loss spectroscopy experiments.
Figure \ref{bs_reflectivity} (c)
shows the calculated loss spectra of Na$_{1.0}$WO$_3$, which
agree well with experiment~\cite{M_Kielwein_PRB_51_10320_1995}.
Both the plasmon and interband excitations can be clearly identified. 

\begin{figure}
\begin{center}
\begin{tabular}{c}
\resizebox{3.0in}{!}{\includegraphics[angle=-90]{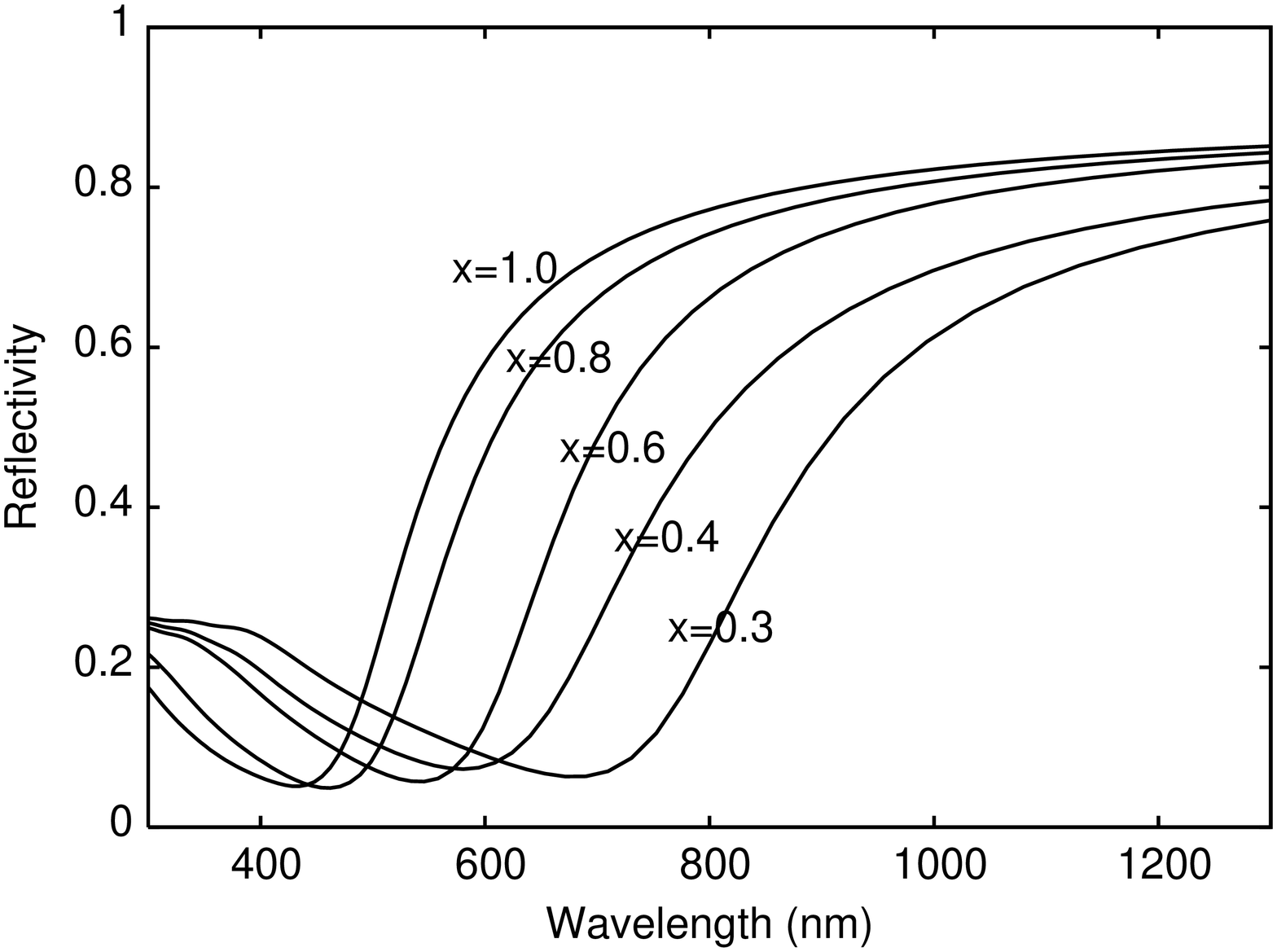}} \\
\resizebox{3.0in}{!}{\includegraphics[angle=0]{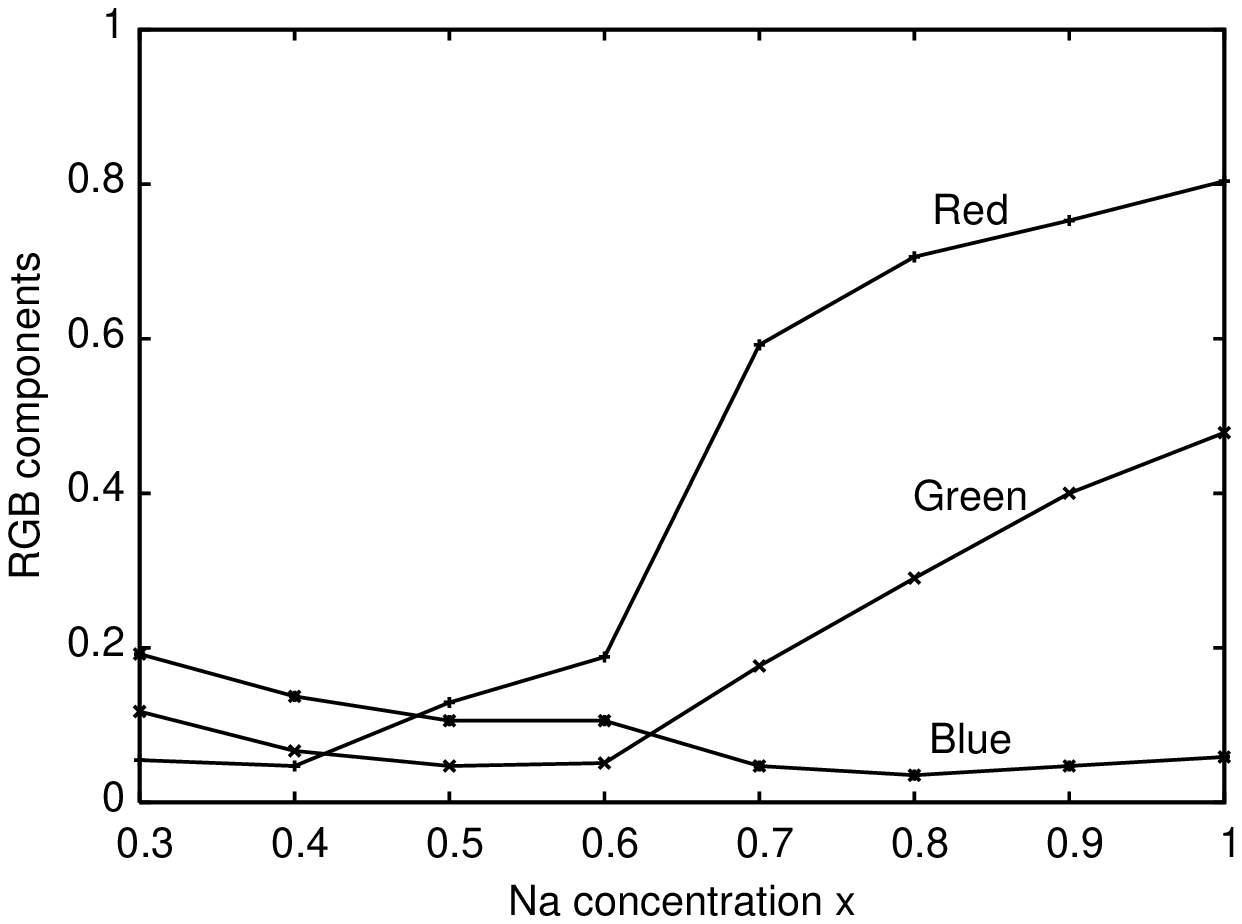}} \\
\resizebox{4.5in}{!}{\includegraphics[angle=0]{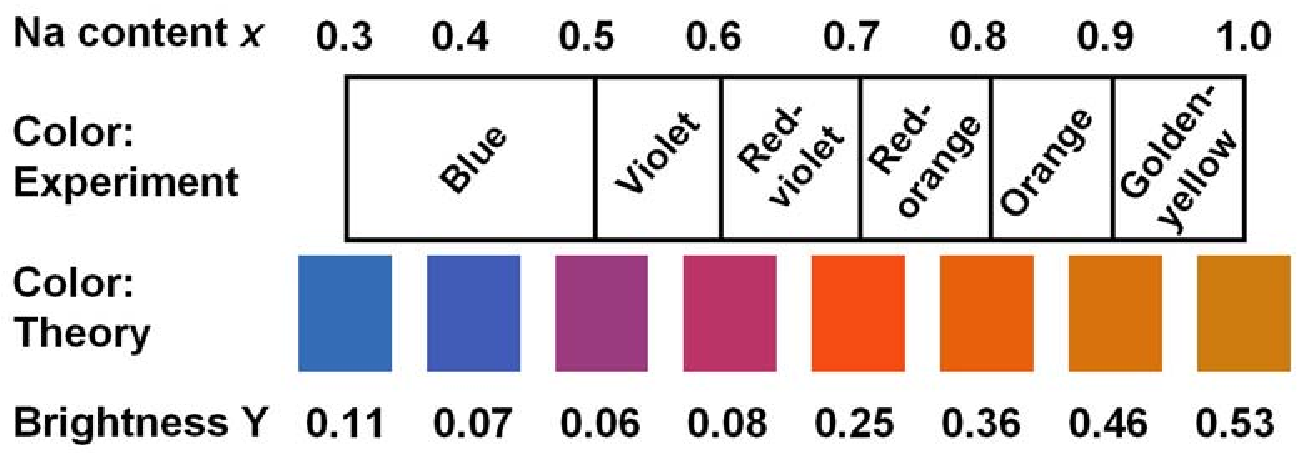}}\\
\end{tabular}
\caption{Calculated reflectivity spectra (top), the RGB tristimulus values (middle),
and the color of Na$_{x}$WO$_3$ under uniform illumination (bottom)}. 
\label{reflectance_rgb_comp}
\end{center}
\end{figure}

{\it Optical properties and the color of Na$_x$WO$_3$ ---}
The results for Na$_{1.0}$WO$_3$ encourages us to carry out a systematic study of the optical 
properties of Na$_x$WO$_3$, with the hope that
a quantitative understanding of dramatic color change of Na$_x$WO$_3$ with varying $x$ might be within the reach of 
today's first-principles electronic structure techniques. Figure \ref{reflectance_rgb_comp} (c) shows the 
calculated reflectivity for Na$_x$WO$_3$ ($0.3\le x \le 1.0$). 
The onset of the free-carriers enhanced reflectivity shifts to shorter wavelengths as $x$ increases.
This is because the screened plasma frequency $\tilde{\omega}_p$ 
increases with increasing free-carrier concentration.
These results agree very well with the measurements by Brown and 
Banks~\cite{Brown_Banks} and by Goldner {\it et al.}~\cite{Goldner} except for
an overall rescaling of the absolute magnitude of the reflectivity (likely due to sample quality and/or diffuse loss of light intensity).

With these results in hand, we are now ready to calculate the 
XYZ and RGB tristimulus values (defined in Eqn.\ \ref{e1} and \ref{e2}) of Na$_x$WO$_3$, 
and hence their colors. We assume that the illumination light source is uniform, i.e.,  $I(\lambda)=$ constant in Eqn.\ \ref{e1}. 
Figure \ref{reflectance_rgb_comp} (b) shows the calculated RGB components as functions of $x$. 
Note that the RGB values are normalized to unity for perfect reflectors [$R(\lambda)=100\%$].
At low Na concentrations ($0.3\le x \le 0.4$), the blue component dominates. This couples with the overall low reflectivity to give a
dark bluish color for these systems as shown in Fig.\ \ref{reflectance_rgb_comp} (c). 
In order to better illustrate the {\it color}, the RGB tristimulus values are rescaled to 
give an even brightness in Fig.\ \ref{reflectance_rgb_comp} (c). 
The calculated values for the relative brightness are also shown in the
figure. The brightness values are normalized such that a perfect reflector
will have a brightness of unity.

As $x$ increases beyond 0.6, both the red and green components are greatly enhanced, 
resulting in the red-orange to golden-yellow metallic appearance of these systems (shown in
Fig.\ \ref{reflectance_rgb_comp} (c). For intermediate Na concentration ($0.5 \le x\le 0.6$),
red and blue components dominate; the systems thus appear violet to red-violet.
Therefore, our theory is able to explain the systematic color change 
of Na$_x$WO$_3$ from dark blue to metallic golden-yellow as $x$ increases from 0.3 to 1.0.  
We mention that the description of color is somewhat subjective and it is
not surprising that slightly different descriptions (e.g., purple v.s. violet)~\cite{handbook,JB_Goodenough,S_Raj}
of the apparent color of Na$_x$WO$_3$ are seen in literature.

In summary, we have carried out first-principles calculations 
of the optical properties, and hence the color, of Na$_x$WO$_3$.
The apparent color is constructed using the CIE 1931
color matching functions and the calculated reflectivity spectra.
Our results clearly explains the systematic color change of Na$_x$WO$_3$ as $x$ varies and
provide direct evidence that the coloration of WO$_3$ upon charge insertion
is a result of the subtle interplay between the interband and intraband optical response.
The reflectivity is greatly enhanced below the screened plasma frequency, which explains
the vivid color change of Na$_x$WO$_3$ as Na concentration increases.
In addition, we demonstrate that first-principles electronic structure
theory can predict the apparent color of crystalline materials and may
help to design materials with desired colors for various applications.


\begin{acknowledgments}
We thank Dr. M. D. Jones for his assistance in coding. This work was supported in part by 
National Science Foundation Grant No. CBET-0844720 and by the UB 2020
Interdisciplinary Research Development Fund (IRDF).
We acknowledge the computational support provided by the 
Center for Computational Research at the University at Buffalo, SUNY.
\end{acknowledgments} 

\newpage

\bibliography{nawo3}
\bibliographystyle{apsrev}

\newpage

\newpage

\newpage

\end{document}